\documentclass[a4paper]{jpconf}
\usepackage{graphicx}
\begin{document}
	\title{Open charm measurements at CERN SPS energies with the new Vertex Detector of the NA61/SHINE experiment}
	
	\author{Anastasia Merzlaya (for NA61/SHINE Collaboration)}
	\address{Jagiellonian University, Krakow, Poland \\
		Saint Petersburg State University, Saint Petersburg, Russia}
	
	\ead{anastasia.merzlaya@cern.ch}

	\begin{abstract}
		The heavy-ion programme of the NA61/SHINE experiment at the CERN SPS has been expanded to allow precise measurements of exotic particles with short lifetime. The study of open charm meson production is a sensitive tool for new detailed investigations of the properties of hot and dense matter formed in nucleus-nucleus collisions. In particular, it offers new possibilities for studies of such phenomena as in-medium parton energy loss and quarkonium dissociation and possible regeneration, thus providing new information to probe deconfinement.
		
		A new high resolution Vertex Detector was designed for the NA61/SHINE experiment for measurements of the very rare processes of open charm production in nucleus-nucleus collisions at the SPS. It will meet the challenges of track registration and of very precise spatial resolution in primary and secondary vertex reconstruction required for the identification of $D^0$ mesons.
		
		A small-acceptance version of the Vertex Detector, the SAVD (Small Acceptance Vertex Detector), was installed last year with a Pb target in the Pb beam of 150$A$ GeV/c momentum, and a modest set of data was collected. The main goal of the ongoing data analysis was to observe a signal from $D^0$ mesons. The status of this analysis will be presented, discussing a number of challenges related to the tracking in the inhomogeneous magnetic field, the matching of SAVD tracks to TPCs tracks, and the extraction of physics results.
	\end{abstract}
	
	\section{Open charm measurements in NA61/SHINE}
	The SPS Heavy Ion and Neutrino Experiment (NA61/SHINE) \cite{NA61} at CERN was designed for studies of the properties of the onset of deconfinement and search for the critical point of strongly interacting matter. 
	These goals are being pursued by investigating p+p, p+A and A+A collisions at different beam momenta from 13$A$ to 158$A$ GeV/c for ions and upto 400 GeV/c for protons.

	Heavy flavour quarks are produced at the early stage of the hadron collision and may serve as sensitive probes for properties of the QGP. However, only indirect measurements of open charm production in nucleus-nucleus collisions at the SPS energies exist, but precise data on both open and hidden charm production are required to provide information on the in-medium behaviour of quarkonia in a model independent way \cite{Satz}.
	
	For open charm measurements in nucleus-nucleus collisions at SPS energies, the experiment NA61/SHINE was upgraded with the new Small Acceptance Vertex Detector (SAVD) \cite{SAVD_Pb-Pb_add, SAVD_sim1, SAVD_sim}. The low yields of $D$ mesons and their short lifetime require detectors located close to the primary vertex that are capable of providing high tracking efficiency and high primary and secondary vertex resolution.
	
	Since $D^0$ mesons typically decay tens to hundreds of microns downstream of their production point, one needs to reconstruct the decay vertex with a precision of about 50$\mu$m in order to recognize daughter particles that originate from  $D^0$ decays.
	
	\begin{figure}[h]
		\begin{center}
			\begin{minipage}[h]{0.5\linewidth}
				\includegraphics[width=9cm,keepaspectratio]{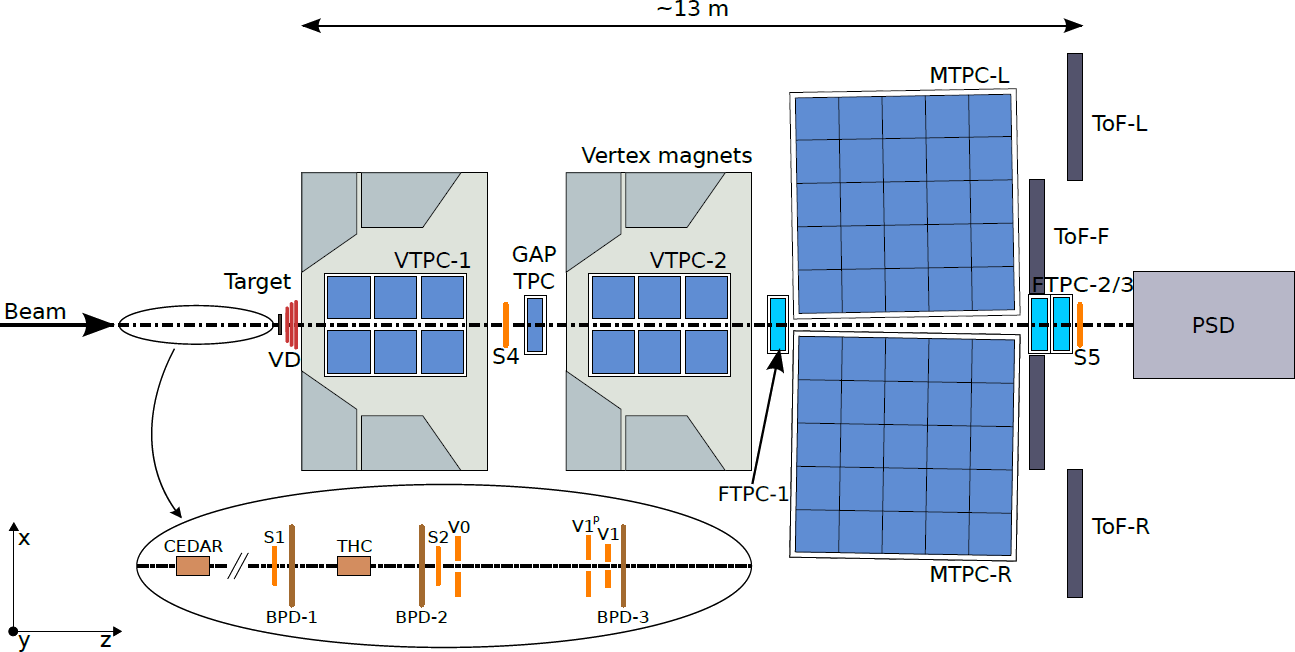}
				\vspace{0.0cm}
				\caption{The layout of the experimental setup of NA61/SHINE.}
				\label{fig:Figure1}
			\end{minipage}
			\hfill 
			\begin{minipage}[h]{0.4\linewidth}
				\includegraphics[width=5.7cm,keepaspectratio]{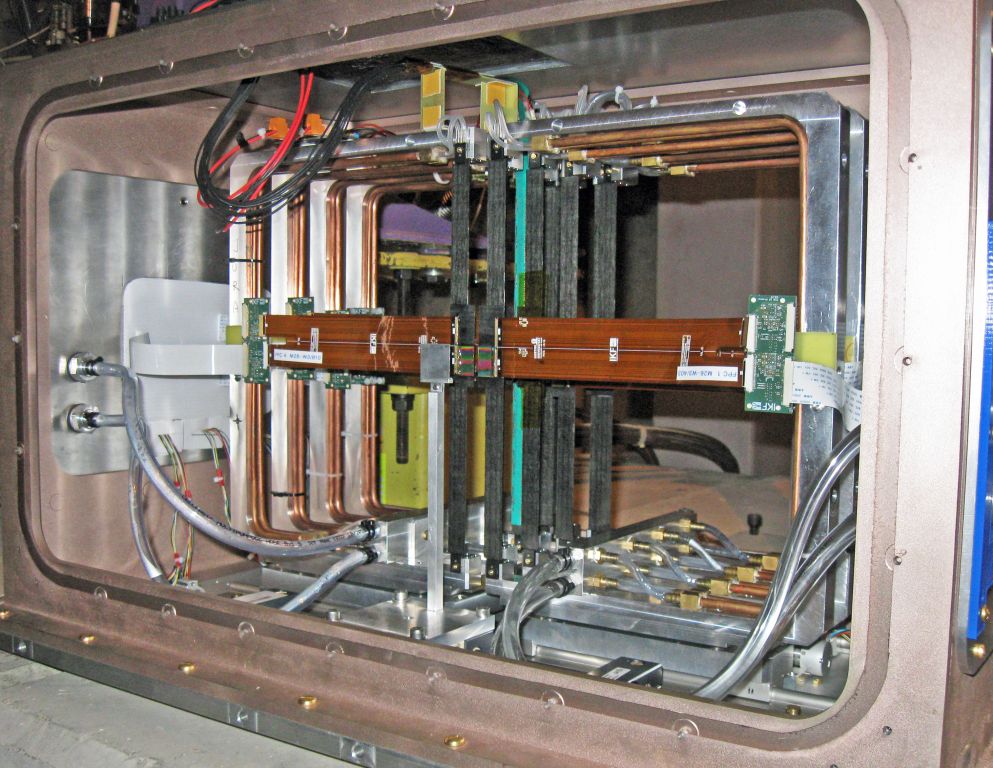}
				\vspace{0.0cm}
				\caption{Photograph of the SAVD before closing the detector with the front and exit windows.}
				\label{fig:Figure2}
			\end{minipage}
		\end{center}
	\end{figure}
	
	\section{Small Acceptance Vertex Detector SAVD for NA61/SHINE}
	The requirements of high precision and resolution can only be reached by adding Vertex Detector to the NA61/SHINE set-up.
	The constructed SAVD is positioned between the target and the VTPC-1 (see Fig.~\ref{fig:Figure1}). It consists of two arms (called Jura and Saleve) with four planes (stations) of coordinate-sensitive detectors located at 5, 10, 15 and 20 cm distance from the target.
	High coordinate resolution MIMOSA-26 sensors \cite{mimosa} based on the CMOS Pixel Sensor technology were chosen as the basic detection element of the SAVD stations. The sensors provide a spatial resolution of 3.5 $\mu$m, have very low material budget (50 $\mu$m thickness), and their readout time is 115.2 $\mu$s. 
	
	In order to reduce beam-gas interactions, the SAVD and the target were placed in a helium enclosure. The magnetic field in the SAVD volume is small and inhomogeneous (0.13 - 0.25T) because the detector is located close to the edge of the VTPC-1 magnet.
	
	~\
	
	Figure~\ref{fig:Figure2} shows the fully integrated detector in the helium vessel installed in December 2016 in the NA61/SHINE experiment for data taking with Pb+Pb collisions at 150$A$~GeV/c~\cite{StatusReport2017}.
	The main goal of the test was to prove precise tracking in the large track multiplicity environment, to demonstrate the ability of precise primary vertex reconstruction and eventually to reconstruct a $D^0$ signal from the collected data.
	
	~\
	\section{Data reconstruction}
	
	\subsection{Track and vertex reconstruction}
	
	A straight line was chosen as the track model: $x(z)=Az+x_0,\hspace{0.5cm} y(z)=Bz+y_0$.
	
	The first step of track reconstruction uses a combinatorial method of track recognition based on checking of combinations of all hits from different stations. If the hits lie on a straight line according to a $\chi^2$ criterion, the combination is accepted as a reconstructed track.
	
	From the distributions of the residuals of hits from the reconstructed tracks the spatial sensor resolution was determined to have values of $\sigma_x$  = 4.7$\mu$m and $\sigma_y$ = 5.0$\mu$m.
	
\begin{figure}[h]
		
		\includegraphics[width=9cm]{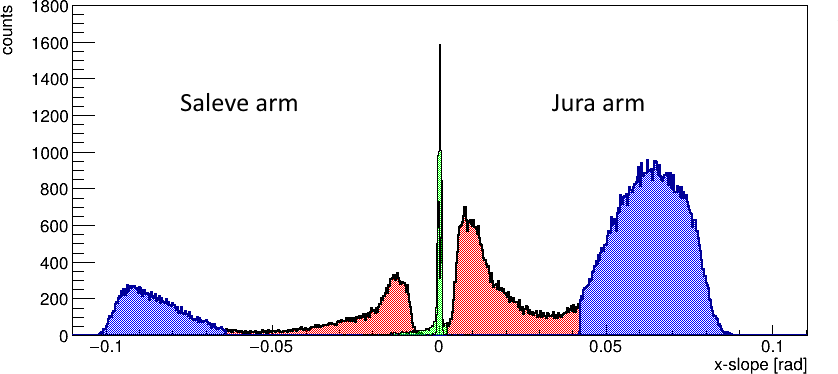}\hspace{2pc}%
		\begin{minipage}[b]{0.37\linewidth}\caption{\label{fig:Figure3} Angular distribution of reconstructed tracks in the $x,z$ plane (x-slope) for Jura (positive values) and Saleve (negative values) arms.}
		\end{minipage}
	\end{figure}

	The angular distribution of the reconstructed tracks in the $x,z$ plane is shown in Fig.~\ref{fig:Figure3}. The distribution for each arm a has clear three-peak structure for each arm. The narrow inner most peak (small angles) was associated with particles produced far upstream and travelling parallel to the beam for a long distance. The middle peak corresponds to particles produced upstream of the target. The outer peak is created by particles produced on the Pb target; these tracks are used for further analysis.
	
	The primary vertex is defined as the point of the closest convergence of all reconstructed tracks. Thus, the longitudinal coordinate of the primary vertex is found by minimizing the expression:
	\begin{eqnarray}
	\nonumber
	D(z)=\sum_{i<j}\lbrace(A_iz+x_i^0-A_jz-x_j^0)^2+(B_iz+y_i^0-B_jz-y_j^0)^2 \rbrace ,
	\end{eqnarray}
	which describes the sum of the relative distances of all track pairs reconstructed in a single event at the given transverse plane defined by the longitudinal coordinate $z$. The $x_{prim}$ and $y_{prim}$ coordinates are afterwards calculated as the average of $x$ and $y$ positions of tracks at $z = z_{prim}$. 
	
\begin{figure}[h]	
		\includegraphics[width=0.7\linewidth]{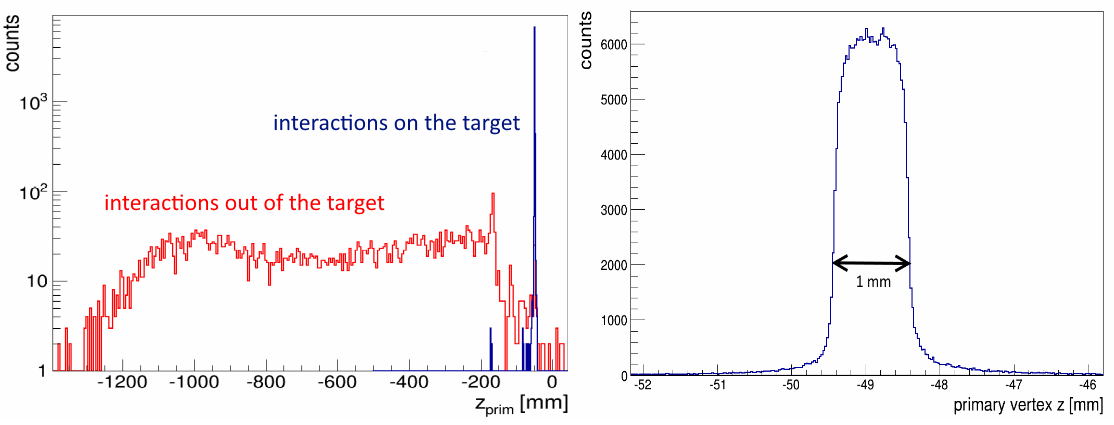}\hspace{2pc}%
		\begin{minipage}[b]{0.23\linewidth}\caption{\label{fig:Figure4} Left: Distribution of the longitudinal coordinate $z_{prim}$ of primary vertices (see text). Right: Distribution of $z_{prim}$ for tracks produced in the target.}
		\end{minipage}
	\end{figure}

	Primary vertex reconstruction was done separately for two samples of tracks with x-slope in the interval 0.02 to 0.04 and with x-slope $>$ 0.04. By looking at the longitudinal distribution of the primary vertex for these samples of tracks (the left picture in Fig.~\ref{fig:Figure4}) one may see that indeed the tracks associated with the most outer peak in Fig.~\ref{fig:Figure3} (blue) originate from the target which is located about 49 mm upstream the fist VD station; the primary vertices associated with tracks from the middle peak (red in Fig.~\ref{fig:Figure3}) are relatively smoothly distributed upstream of the target in the range from -1200 mm to -55 mm. At -190 mm the distribution has a sharp peak which is related to interactions in the aluminized Mylar front window of the SAVD. One can also see that between the window and the target the frequency of interaction drops due to the presence of helium gas in the SAVD vessel. 
	
	The distribution of $z_{prim}$ from the interactions in the target is shown by the right panel of Fig.~\ref{fig:Figure4}. The near rectangular shape of the distribution allows to reconstruct the thickness of the target (= 1 mm) and its location with respect to the SAVD. The primary vertex resolution  is  $\sigma_x$ = 5$\mu$m, $\sigma_y$ = 1.5$\mu$m and $\sigma_z$ = 30$\mu$m (for more details see \cite{StatusReport2017}). The difference between  $\sigma_x$ and $\sigma_y$ is caused by the presence of a vertical component of the magnetic field in the SAVD volume.
	
	After the primary vertex is found, the next step of track reconstruction searches for 3-hit tracks using the Hough transform method (for more details see \cite{Merzlaya2017}). It uses a parametric description of the track by a set of parameters.
	Once a parametric description of the track in the coordinate space is given, the hit position measurements are transformed into the track parameter space.
	Then, in this so-called accumulation space track candidates are obtained as local maxima using the voting procedure: the most popular candidates are accepted as real tracks. 
	
	~\
	
	The magnetic field in the Vertex Detector volume allows momentum reconstruction of the SAVD-tracks \cite{Merzlaya2017}. The following formula was used to calculate  momentum:
	\begin{eqnarray}
	\nonumber
	p_{xz}=\frac{Ze\int B_ydl}{\sin\alpha _1 - \sin\alpha _2},
	\end{eqnarray}
	where $\alpha _1$, $\alpha _2$ are angles of the track in the first and last stations respectively. Momentum reconstruction is imprecise due to the fact that the field is weak and inhomogeneous ($\int B_ydl$ = 0.04 T$\cdot$m).  Nevertheless, the obtained values are useful for verifying track matching between SAVD-tracks and TPC-tracks.
	
	\subsection{Track matching}
	
	The track matching algorithm tries all possible combinations of pairs of SAVD with TPC tracks, assigning to each SAVD track the total momentum of the paired TPC track (which is precise). 
	Next, the SAVD tracks and TPC tracks are extrapolated to a common surface. The algorithm then searches for the best match among all combinations based on the difference in the position and direction of the tracks and uses charge and total momentum to verify the match (since the SAVD has small reconstruction power for charge and momentum).
	About 80\% of SAVD tracks are in the acceptance of the TPCs according to GEANT4 simulations. The track matching efficiency is determined by number of matched SAVD tracks divided by number of SAVD tracks in the TPC acceptance, and is on the level of 90\%.
	
	~\
	\section{Reconstruction of $D^0$ signal}
	
	SAVD tracks matched to TPC tracks are used to search for the $D^0$ signal. In the current analysis PID information was not used. Each SAVD track is paired
	with another SAVD track and is assumed to be either a kaon or pion. Thus each pair contributes twice in the combinatorial invariant mass distribution.
	The combinatorial background is several orders of magnitude higher than the $D^0 + \bar{D^0}$ signal due to the low yield of charm particles.
	In order to reduce the large background and maximize the signal to noise ratio (SNR) of the reconstructed $D^0$ peak, four cuts are applied, which were determined from the simulations \cite{SAVD_sim1, SAVD_sim}:
	\begin{itemize}
		\item A cut on the track transverse momentum $p_T>0.34$ GeV/c;
		\item A cut on the track impact parameter $d>34$ $\mu$m;
		\item A cut on the longitudinal distance between the $D^0$ decay candidate and the interaction point $V_z > 475$ $\mu$m;
		\item A cut on the impact parameter D of the back-extrapolated $D^0$ candidate momentum vector $D < 21$ $\mu$m.
	\end{itemize}
	
\begin{figure}[h]	
		\includegraphics[width=0.7\linewidth]{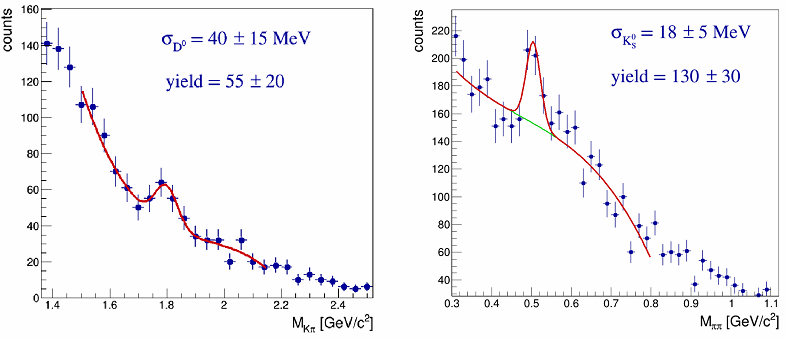}\hspace{2pc}%
		\begin{minipage}[b]{0.25\linewidth}\caption{\label{fig:Figure5}Left: Invariant mass distribution of unlike charge $\pi, K$ decay track candidates. Right: Invariant mass distribution of $\pi^-, \pi^+$ decay track candidates. See text for details.}
		\end{minipage}
	\end{figure}

	The left panel of Fig.~\ref{fig:Figure5} shows the invariant mass distribution of unlike charge daughter candidates with applied cuts. One observes a small peak emerging at 1.8 GeV/$c^2$, which could be an indication of $D^0$ production.
	The invariant mass distribution was fitted (red line) using of a third order polynomial function to describe the background and a Gaussian peak for the $D^0$ signal contribution. The indicated errors are statistical only. From the fit one finds the width of the peak to be 40 $\pm$ 15 MeV and the total yield to amount to 55 $\pm$ 20 with a SNR of 3.  
	
	To test the detector capabilities and validate the $D^0$ peak extraction procedure, it was also attempted to reconstruct the $K^0_s$, which is much more abundantly produced. However, one can reconstruct only a small fraction of all $K^0_s$ with the SAVD since the detector design and the tracking algorithm were optimized for decays which have much shorter decay length. 
	A clear peak corresponding to $K^0_s$ decays is seen in the right panel of Fig.~\ref{fig:Figure5} with a width of 18 $\pm$ 5 MeV and a total yield of 130 $\pm$ 30 with a SNR of about 4.5.
	
	\section{Summary}
	
	A small-acceptance version of the NA61/SHINE Vertex Detector (SAVD) was installed in 2016. A modest set of data of Pb+Pb collisions at 150$A$~GeV/c was collected and analysed. A weak indication of a $D^0$ signal was observed in the $\pi, K$ decay channel. However, the data set is not finally calibrated, and further optimization of tracking, track matching and analysis algorithms, as well as of cuts is still ongoing. It may be premature to draw a final conclusion from the analysis result, though the result appears promising.
	
	The SAVD  will be used in p+Pb and Xe+La data taking in the autumn of 2017. Also, high statistics measurements in Pb-Pb collisins at 150$A$~GeV/c are planned in 2018.
	Looking forward, an upgraded version the so-called Large Acceptance Vertex Detector (LAVD) with more sensors is being planned; the exact design of this detector is currently under investigation.

	\section*{Acknowledgements}
	The work related to instrumentation was supported by the Polish National Center for Science grant 2014\slash15\slash B\slash ST2\slash02537 and the work related to reconstruction and data analysis was partially supported by the Russian Science Foundation research grant 16-12-10176.
	
\end{document}